\documentclass[useAMS,usenatbib,a4paper]{mn2e}
\usepackage[dvips]{graphicx}
\usepackage{hyperref}
\usepackage{epstopdf}
\usepackage{times}
\usepackage{amsmath}
\usepackage{amssymb}
\usepackage{color}
\newcommand{\aap}{A\&A}

\newcommand{\mnras}{MNRAS}

\newcommand{\apj}{ApJ}
\newcommand{\apjl}{ApJ}
\newcommand{\apjs}{ApJS}
\newcommand{\pasj}{PASJ}

\newcommand{\aj}{AJ}

\newcommand{\prd}{Phys. Rev. D}
\newcommand{\prl}{Phys. Rev. Lett.}
\hypersetup{
  colorlinks,
  citecolor=blue,
  linkcolor=blue,
  urlcolor=blue
}
\title[Black Hole Mergers from Globular Clusters]{Binary Black Hole Mergers from Globular Clusters: the Impact of Globular Cluster Properties}
\author[J. Hong et al.]{Jongsuk Hong$^{1,2}$\thanks{E-mail: jongsuk.hong@pku.edu.cn (JH)}, Enrico Vesperini$^2$, Abbas Askar$^{3,4}$, Mirek Giersz$^3$, \newauthor Magdalena Szkudlarek$^5$ and Tomasz Bulik$^6$ 
\\
  $^1$ Kavli Institute for Astronomy and Astrophysics, Peking University, Yi He Yuan Lu 5, HaiDian District, Beijing 100871, China\\
  $^2$ Department of Astronomy, Indiana University, Bloomington, Swain West, 727 E. 3rd Street, IN, 47401, USA\\
  $^3$ Nicolaus Copernicus Astronomical Centre, Polish Academy of Sciences, ul. Bartycka 18, 00-716 Warsaw, Poland\\
  $^4$ Lund Observatory, Department of Astronomy, and Theoretical Physics, Lund University, Box 43, SE-221 00 Lund, Sweden\\
  $^5$ Janusz Gil Institute of Astronomy, University of Zielona G\'ora, Licealna 9, 65-407 Zielona G\'ora, Poland\\
  $^6$ Astronomical Observatory, University of Warsaw, Al. Ujazdowskie 4, 00-478 Warsaw, Poland\\
}
\begin{document}

\date{Accepted 2018 August 10. Received 2018 July 11; in original form 2018 May 10}
\maketitle

\label{firstpage}

\begin{abstract}
The dense environment of globular clusters (GCs) can facilitate the formation of binary black holes (BBHs), some of which can merge with gravitational waves (GW) within the age of the Universe. We have performed a survey of Monte-Carlo simulations following the dynamical evolution of GCs with different masses, sizes and binary fractions and explored the impact of the host GC properties on the formation of BBH mergers. We find that the number of BBH mergers from GCs is determined by the GC's  initial mass, size and primordial binary fraction. We identify two groups of BBH mergers: a primordial group whose formation does not depend on cluster's dynamics and a dynamical group whose formation is driven by the cluster's dynamical evolution. We show how the BBH origin affects the BBH mergers' main properties such as the chirp mass and merging time distributions.
We provide analytic expressions for the dependence of the number of BBH mergers from individual GCs on the main cluster's structural properties and the time evolution of the merger rates of these BBHs. These expressions provide an essential ingredient for a general framework allowing to estimate the merger rate density.
Using the relations found in our study, we find a local merger rate density of 0.18--1.8 ${\rm Gpc}^{-3}{\rm yr}^{-1}$ for primordial BBH mergers and 0.6--18 ${\rm Gpc}^{-3}{\rm yr}^{-1}$ for dynamical BBH mergers, depending on the GC mass and size distributions, initial binary fraction and the number density of GCs in the Universe. 
\end{abstract}

\begin{keywords}
globular clusters: general --- stars: black holes --- gravitational waves 
\end{keywords}

\section{Introduction\label{S1}}
Recently, Advanced LIGO has made the first detection of gravitational waves (GWs) and opened a new window to explore very energetic events \citep{2016PhRvL.116f1102A}. The event responsible for the GWs revealed by this first detection, GW150914, was the merger of black holes (BHs) in a binary system and it has been followed by four more detections of merging binary BHs \citep[BBHs,][]{2016PhRvL.116x1103A,2016PhRvX...6d1015A,2017PhRvL.118v1101A,2017PhRvL.119n1101A} and one merging binary neutron stars \citep{2017PhRvL.119p1101A}.

A number of different scenarios for the formation of these merging compact binaries have been proposed so far; the different formation mechanisms proposed have invoked isolated binary evolution
\citep[e.g.][]{2002ApJ...572..407B,2007ApJ...662..504B,2012ApJ...759...52D}, 
three-body interactions in dense stellar systems \citep[e.g.][]{2000ApJ...528L..17P,2006ApJ...637..937O,2016ApJ...824L...8R,2018ApJ...855..124S}, 
the orbital evolution of hierarchical systems \citep[e.g.][]{2012MNRAS.422..841A,2014ApJ...781...45A,2016ApJ...816...65A,2018ApJ...856..140H,2018arXiv180508212R}, 
relativistic captures \citep[e.g.][]{2009MNRAS.395.2127O,2015MNRAS.448..754H,2017PhRvD..96h4009B,2018ApJ...860....5G}. 
As for the environment in which these compact binaries might form, the scenarios proposed include globular clusters
\citep[GCs,][]{2010MNRAS.402..371B,2011MNRAS.416..133D,2013MNRAS.435.1358T,2014MNRAS.440.2714B,2015PhRvL.115e1101R,2016PhRvD..93h4029R,2017ApJ...834...68C,2017PASJ...69...94F,2017MNRAS.464L..36A,2017MNRAS.469.4665P,2017arXiv170607053B,2018A&A...615A..91B,2018PhRvD..97j3014S}, young/open clusters \citep[e.g.][]{2014MNRAS.441.3703Z,2017MNRAS.467..524B,2018MNRAS.473..909B} and galactic nuclei \citep[e.g.][]{2009MNRAS.395.2127O,2012ApJ...757...27A,2018MNRAS.474.5672L}.

An important aspect concerning the formation of BHs is the mass fall-back after the supernova explosions. As discussed by \citet{2002ApJ...572..407B}, this fall-back (i.e., failed supernovae) mechanism can increase the remnant BH masses and reduce the natal kicks, which, in turn, can lead to a larger fraction of BHs retained inside the host stellar system  \citep{2015ApJ...800....9M,2016MNRAS.458.1450W, 2016MNRAS.463.2109R,2018MNRAS.478.1844A,2018MNRAS.479.4652A}. 

The retention of a large number of BHs can significantly influence not only the internal dynamics \citep[e.g.][]{2013MNRAS.432.2779B} but also the observational properties of star clusters \citep[e.g.][]{2008MNRAS.386...65M,2017ApJ...834...68C,2017arXiv171203979W,2018ApJ...855L..15K,2018MNRAS.476.5274L,2018MNRAS.479.4652A,2018MNRAS.478.1844A}. The retained BHs in a dense stellar system rapidly sink to the centre of the system due to the effects of dynamical friction and form a compact subsystem predominantly composed of BHs, on a timescale of few hundreds Myr \citep{2013ApJ...763L..15M}. Due to its short relaxation timescale, a BH subsystem quickly undergoes core collapse and generate energy through the formation and dynamical interactions of BBHs \citep{2013MNRAS.432.2779B}. Recoil velocities acquired during binary-single and binary-binary interactions can result in BHs ejection from GCs, and some numerical studies \citep[e.g.][]{2015ApJ...800....9M,2017MNRAS.469.4665P} suggested that $\sim$30\% of dynamically escaping BHs are in binary systems, some of which are expected to merge within the age of the Universe. Moreover, \citet{2018MNRAS.478.1844A} suggested that some massive Galactic GCs (GGCs) are still harbouring a large number of BHs and that the formation and ejection of BBHs can still be ongoing in those GGCs. The BBHs' properties as well as the merger and detection rates of these BBHs are significantly affected not only by the global properties of host GCs such as the initial mass, size and the metallicity  \citep[e.g.][]{2016PhRvD..93h4029R,2017MNRAS.464L..36A,2017ApJ...836L..26C} but also by the GC's stellar initial mass function and the prescriptions for the mass fall-back and the stellar wind \citep[see e.g.][and the references therein]{2017ApJ...834...68C}.

In this paper, we present an analysis of the survey of Monte-Carlo simulations of GCs evolution from \citet{2017MNRAS.464.2511H} and of another set of simulations performed specifically for this paper aimed at a detailed characterization of the link between the properties of BBHs formed in GCs and the structure of the host GCs. Understanding the connection between the properties of the BBHs and those of their host GCs is an important step for more realistic estimates of the merger rate of BBHs. We extracted the information of all escaping BBHs from our GC simulations and found some empirical relations between the properties of merging BBHs and those of the host GCs. These relations provide an essential ingredient to estimate the merger and detection rates of BBHs for any assumed GC system properties and GC formation rate. We also provided examples of estimates of the local merger rate density for various assumptions concerning the properties of GC systems.

The structure of this paper is as follows. We briefly describe the numerical method, the initial conditions and assumptions of our GC simulations in Section \ref{S2}. The relations between the properties of merging BBHs and those of the host GC  are presented in Section \ref{S3}. In Section \ref{S4}, we then estimate the local merger rate density based on these relations. We conclude with a summary of our results in Section \ref{S5}.

\section{Methods and Initial Conditions\label{S2}}
The models used for this study are those of the survey of Monte-Carlo simulations presented in \citet{2017MNRAS.464.2511H}.
The simulations followed the evolution of 81 cluster models with a variety of initial number of stars ($N=2\times10^5, 5\times10^5, 10^6$), half-mass radii ($r_{\rm h}=1, 2, 4$ pc), binary fractions (10, 20, 50 per cent) and galactocentric distances ($r_{\rm G}=4, 8, 16$ kpc) \citep[see Table 1 in ][]{2017MNRAS.464.2511H} and were run with the {\sc mocca} code \citep{2013MNRAS.431.2184G,2013MNRAS.429.1221H}. The initial density structure of clusters follows the \citet{1966AJ.....71...64K} density profile with the dimensionless central potential, $W_{0} = 7$. We have adopted \citet{2001MNRAS.322..231K} initial mass function with the mass range of stars from 0.1 to 100 M$_{\odot}$.  The metallicity is fixed to $Z = 0.001$ for all our simulation models. All single and binary stars in the simulations evolve according to the stellar evolution recipes \citep[SSE \& BSE,][]{2000MNRAS.315..543H,2002MNRAS.329..897H} implemented in the {\sc mocca} code. We used the stellar wind prescription of SSE and BSE. In all of our simulations, we adopt the mass fall-back mechanism \citep{2002ApJ...572..407B} modifying the natal kicks for BHs. 

All our GC simulation models are limited by the tidal field from the host galaxies with a realistic treatment of escaping stars based on  \citet{2000MNRAS.318..753F} \citep[see also][]{2013MNRAS.431.2184G}. For the parameter spaces of the initial conditions considered in \citet{2017MNRAS.464.2511H}, the ratio of the half-mass radius to the tidal radius for the GC simulation models ranges from 0.005 to 0.09. 

In \citet{2017MNRAS.464.2511H}, we used the initial binary distribution (e.g. eccentricity, semi-major axis and the mass ratio) based on the {\it initial binary population} (hereafter IBP) in which the orbital parameters of short-period proto-binaries are redistributed by mutual interactions between binary components (e.g. mass transfer and tidal circularization) due to the large stellar radii during pre-main-sequence stage as suggested by \citet{1995MNRAS.277.1507K} \citep[see also ][]{2013pss5.book..115K}.\footnote{This {\it pre-main-sequence eigenevolution} \citep{1995MNRAS.277.1507K} was originally postulated to explain the observed properties of Galactic field binary populations originating in embedded clusters. Most recently, \citet{2017MNRAS.471.2812B,2018MNRAS.474.3740B} have provided a modified prescription of the Kroupa IBP for the binary distributions in GCs, which is, however, not applied in this study.} As the results of this proto-binary evolution, short period binaries with large eccentricity are preferentially depleted and tend to have similar masses.
For this study, we have also run another set of 81 simulations with the {\it birth binary population} (hereafter BBP) from \citet{1995MNRAS.277.1507K} that follows the \citet{1991A&A...248..485D} period distribution and the thermal eccentricity distribution to investigate the effects of the initial orbital properties of primordial binaries on the formation, dynamical evolution of BBHs and the rate of merger events among these BBHs from GCs.

\section{bbh mergers from gc simulation models\label{S3}}
\subsection{Correlation between the number of BBH mergers and GC properties\label{S3.1}}
\begin{figure}
  \centering
  \includegraphics[trim=15 10 5 5,width=1.0\columnwidth]{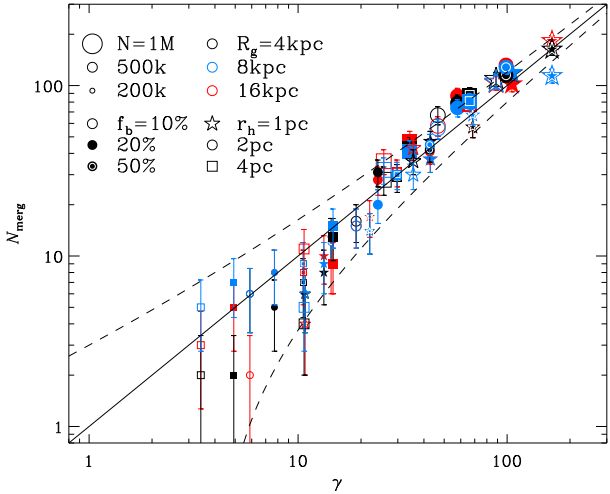}
  \caption{Number of BBH mergers versus the parameter $\gamma$ defined in Eq. (\ref{E1}). Different symbol types, sizes and colors represent models with different initial number of stars, half-mass radii, galactocentric distances and binary fractions \citep[see also][]{2017MNRAS.464.2511H}. Dashed lines indicate the two times of Poisson errors of the locus line.}\label{F1}
\end{figure}
\begin{figure}
  \centering
  \includegraphics[trim=15 10 5 5,width=1.0\columnwidth]{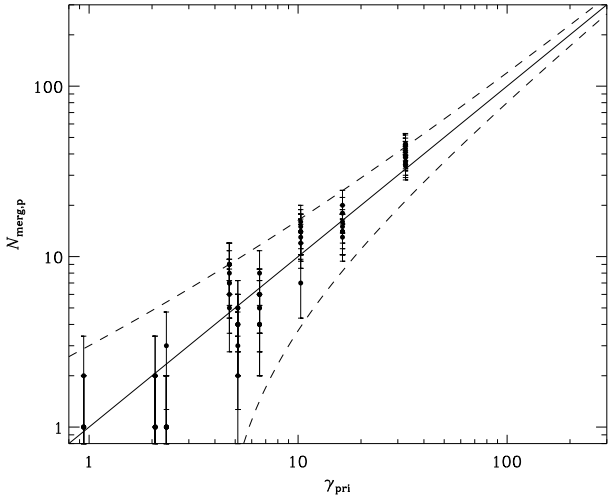}
  \caption{Same as Fig. \ref{F1} but for only primordial BBH mergers that escape from GCs due to the natal kicks during supernova explosions.}\label{F2}
\end{figure}
\begin{figure}
  \centering
  \includegraphics[trim=15 10 5 5,width=1.0\columnwidth]{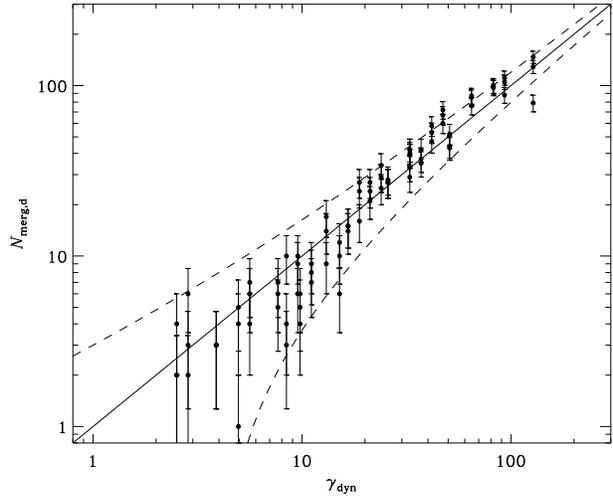}
  \caption{Same as Fig. \ref{F1} but for only dynamical BBH mergers that form dynamically inside GCs by three-body or exchange encounters and subsequently escape.}\label{F3}
\end{figure}
We first focus on the presentation of our results for the simulations with the IBP distribution.
After 12 Gyr of evolution, the 81 GC models explored for this study produced 9519 escaping BBHs and 3402 of them emit GWs and merge within 12 Gyr. 
To illustrate the dependence of the GW events on the properties of the host GCs in which the BBH formed, we show the correlation between the number of merging BBHs, $N_{\rm merg}$ and the initial properties of GCs in Fig. \ref{F1}. 
We determine the number of BBH mergers that escape from GCs and subsequently merge within 12 Gyr, i.e., $t_{\rm merg}\equiv t_{\rm esc}+t_{\rm Peters} <$ 12 Gyr, where $t_{\rm esc}$ and $t_{\rm Peters}$ are, respectively, the BBH escaping time and the \citet{1964PhRv..136.1224P} timescale for GW coalescence of BBHs calculated using their semi-major axis and eccentricity at the moment of escape. We found that the number of merging BBHs is closely correlated with a parameter, $\gamma$, defined as
\begin{equation}\label{E1}
\gamma\equiv A\cdot \frac{M_{0}}{10^{5}{\rm M}_{\odot}}\times \Big(\frac{\rho_{\rm h}}{10^{5}{\rm M}_{\odot}{\rm pc}^{-3}}\Big)^{\alpha} + B\cdot \frac{M_{0}}{10^{5}{\rm M}_{\odot}}\times f_{\rm b,0}    
\end{equation}
where $M_{0}$, $\rho_{\rm h}$ and $f_{\rm b,0}$ are, respectively, the initial total mass, initial half-mass density (i.e., mean density within the half-mass radius) and the initial primordial binary fraction; $A$, $B$ and $\alpha$ are the fitting parameters. Our best fitting result that minimizes the $\chi^{2}$ value is ($A$, $B$, $\alpha$) $=$ (12.53$\pm$0.22, 6.89$\pm$0.84, 0.33$\pm$0.02). The uncertainty on the best-fit parameters is determined by calculating the values for which $\chi^{2}=\chi_{\rm min}^{2}+1$ \citep{1976ApJ...210..642A} by assuming that the $\chi^{2}$ distribution in 1D parameter space is simply a quadratic function.
Eq. (\ref{E1}) implies that there are two main formation channels for BBH mergers from GCs; one, the primordial channel, is related only to the primordial binary fraction and binary stellar evolution and is described by the second term of Eq. (\ref{E1}). The other channel is affected by the cluster's internal dynamics and its contribution to the total number of merging BBHs is described by the first term in Eq. (\ref{E1}) (hereafter we will refer to this as the dynamical channel). The number of BBH mergers from the primordial channel, $N_{\rm merg,p}$, depends, as was to be expected, only on the initial binary fraction and the total mass of GCs. The number of merging BBHs from the dynamical channel, $N_{\rm merg,d}$, on the other hand is the result of the combined effects of a number of processes affected by a variety of structural parameters (e.g. encounter rates, hardening rate per encounters, central velocity dispersion, ejection rate, etc.); our results show that the number of merging BBHs resulting from the complex interplay of all these processes is well described by a parameter with a simple dependence on the cluster's mass and half-mass density. 

We point out the number of merging BBHs does not show any significant dependence on the galactocentric distance. For the compact clusters explored in our survey, this is to be expected as the population of merging BBHs escape from clusters as a result of either natal kicks following supernova explosion or ejection from close encounters in the cluster's inner regions. The galactocentric distance and the strength of the tidal field, on the other hand, are relevant for the more gradual evaporation process which is not important for the BBH population studied here. Note, however, that there is a broad trend of metallicity with galactocentric distance \citep[e.g.][]{1994AJ....108.1292D} that can affect the properties of BH populations and the formation of BBHs accordingly.

In order to better illustrate the relative importance of the two channels, we have divided the merging BBHs escaping from our GC simulation models into two groups according to their origin. We have classified BBHs that escape through the natal kick after supernova explosions  as merger candidates with the primordial origin (789 of all merging BBHs); all the other merging BBHs are classified as dynamical BBH mergers (2613 of all merging BBHs).
Figs. \ref{F2} and \ref{F3} show the dependence of the number of primordial and dynamical  BBH mergers on, respectively,  $\gamma_{\rm pri}$ and $\gamma_{\rm dyn}$, the two terms already introduced in the definition of the parameter $\gamma$ in Eq. (\ref{F1}),
\begin{equation}\label{E2}
\gamma_{\rm dyn}\equiv A\cdot \frac{M_{0}}{10^{5}{\rm M}_{\odot}}\times \Big(\frac{\rho_{\rm h}}{10^{5}{\rm M}_{\odot}{\rm pc}^{-3}}\Big)^{\alpha},
\end{equation}
\begin{equation}\label{E3}
\gamma_{\rm pri}\equiv B\cdot \frac{M_{0}}{10^{5}{\rm M}_{\odot}}\times f_{\rm b,0}.
\end{equation}
The best fitting parameters are ($A$, $\alpha$) $=$ (12.30$\pm$0.44, 0.33$\pm$0.01) for dynamical BBH mergers and ($B$) $=$ (6.64$\pm$0.25) for primordial BBH mergers, separately. It is apparent that the best-fit parameters calculated for the two merging BBH population separately are very similar to those obtained from  fitting all the BBH mergers together. 

\begin{figure}
  \centering
  \includegraphics[trim=15 10 5 5,width=1.0\columnwidth]{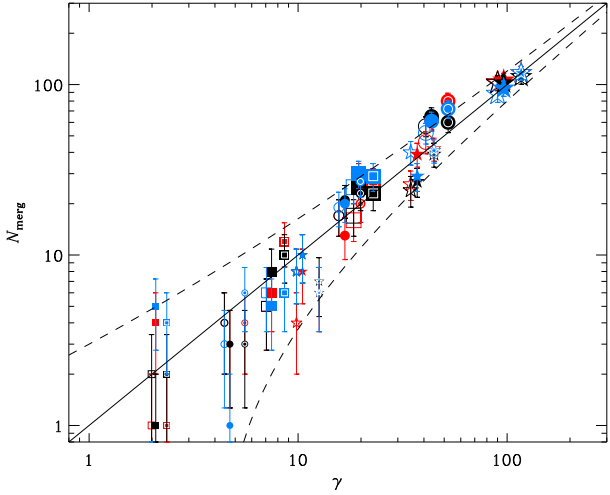}
  \caption{Same as Fig. \ref{F1} but for GC simulation models with the BBP distribution.}\label{F4}
\end{figure}
\begin{figure}
  \centering
  \includegraphics[trim=15 10 5 5,width=1.0\columnwidth]{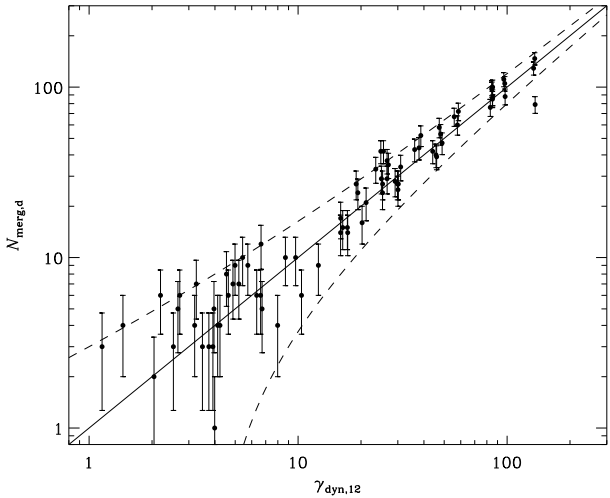}
  \caption{Same as Fig. \ref{F3} but $\gamma_{\rm dyn, 12}$ is defined by using current GC properties.}\label{F5}
\end{figure}
\begin{figure}
  \centering
  \includegraphics[trim=15 10 5 5,width=1.0\columnwidth]{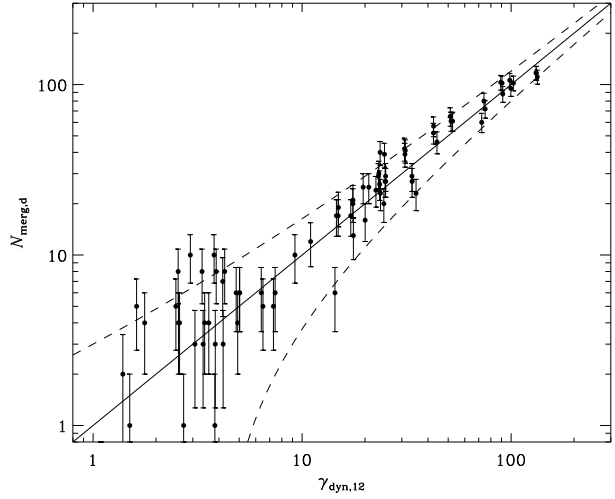}
  \caption{Same as Fig. \ref{F5} but for the BBP distribution.}\label{F6}
\end{figure}
For models with the BBP distribution, we find 6755 BBHs escaping from GCs in total and 2382 of them merge before 12 Gyr. In Fig. \ref{F4}, we show the relation between $N_{\rm merg}$ and $\gamma$ for the simulations with the BBP distribution. The best fitting parameters for this set of simulations are ($A$, $B$, $\alpha$) $=$ (14.55$\pm$0.31, -0.03$\pm$0.03, 0.38$\pm$0.01). For the BBP distribution, we obtained only 1 primordial BBH merger while all the other BBH mergers have a dynamical origin and this explains the lack of dependence of the number of merging BBHs on the initial binary fraction.
This is due to the pairing of primordial binary components of massive stars. For the BBP distribution, we used a uniform pairing so that the masses of secondary stars are uniformly chosen in between  [$m_{\rm min}$, $m_{\rm pri}$]. For the IBP distribution with mass feeding algorithm \citep{2013pss5.book..115K} used in the simulations, the pairing rule for the massive binaries is different.
Some theoretical studies \citep[e.g.][]{2000MNRAS.314...33B,2002MNRAS.336..705B,2007ApJ...661.1034K} for the formation of massive binaries in star forming regions show that massive binaries tend to evolve to the mass ratio $q\sim1$ during the proto-binary stage \citep[note that, however, some mechanisms such as the magnetic breaking during proto-binary evolution can prevent the evolution of mass ratio of proto-binaries toward $q \sim 1$; see e.g.][]{2013ApJ...763....7Z}. 
GC simulation models with IBP distribution in this study adopted this condition so that the massive binaries (especially for $m>5$M$_{\odot}$) that can be the progenitors of BHs are more likely to have $q\sim1$ and evolve to BBHs.\footnote{However, \citet{2017MNRAS.471.2812B,2018MNRAS.474.3740B} provided a modified prescription of the Kroupa IBP distribution for GC environments suggesting that the pre-main-sequence eigenevolution and mass feeding algorithm are not applied to massive binaries and that the pairing rule for massive binaries is a uniform pairing based on \citet{2012Sci...337..444S}. If this is the case, the number of primordial BBH mergers from GCs will be negligible and the relation between the number of BBH mergers and GC properties will be similar to that for the BBP distribution in our study.} On the other hand, with the BBP distribution, massive stars are initially coupled with less massive stars and therefore require exchange encounters to become BBHs. 

Figs. \ref{F5} and \ref{F6} show the correlation between $N_{\rm merg}$ and the current (at $T = 12$ Gyr) GC properties for models with different initial binary distributions. 
We only present the correlation for dynamical BBH mergers in this figure.
The best fitting parameters of Eq. (\ref{E2}) for the correlation between the number of dynamical BBH mergers and the current GC properties are ($A$, $\alpha$) $=$ (303.1$\pm$6.1, 0.43$\pm$0.02) for the IBP distribution and ($A$, $\alpha$) $=$ (527.7$\pm$11.1, 0.52$\pm$0.02) for the BBP distribution, respectively. This relation can be used to estimate current merger rate for the MW GCs or Local Group GCs (see Section \ref{S4.3}). The values of $A$ and $B$ in the relation with the current GC properties are larger than those obtained when the initial GC properties are used because the current masses of GCs are smaller than the initial ones.

\subsection{Time evolution of merger rates\label{S3.2}}
\begin{figure}
  \centering
  \includegraphics[trim=15 10 5 5,width=1.0\columnwidth]{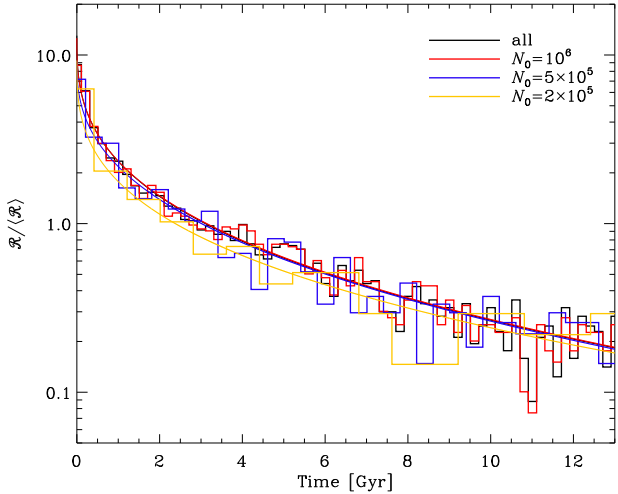}
  \caption{Time evolution of BBH merger rate normalized to the average merger rate over 12 Gyr. Different colors for histograms represent the time evolution of merger rates from GC simulation models with different initial number of stars. Solid lines show the best fitting results (see Section \ref{S3.2} for further details). }\label{F7}
\end{figure}
In the previous section, we provided a relation between the (initial or current) GC properties and the expected number of BBHs that escape from GCs and merge by emitting GWs within 12 Gyr. However, many numerical studies aimed at the estimation of the merger rate of BBHs from GCs have shown that the merger rate is time-dependent \citep[e.g.][]{2017MNRAS.464L..36A,2017PASJ...69...94F}. Since one of our main goals is to provide an empirical relation that allows to calculate the merger rates and the detection rates of BBH merger events from GCs, we have also calculated a model for the time dependence of the rate of BBH mergers from GCs. In Fig. \ref{F7}, we present the histograms of the merging time, $t_{\rm merg}$, of BBHs, with the numbers normalized to the merger rate averaged over the 12 Gyr of evolution. This figure clearly shows that the merger rate decreases with time very rapidly. Initially the merger rate is as high as 10 times the average merger rate while the merger rate at 12 Gyr is $\sim$5 times lower than the average merger rate. We found that the time evolution of the merger rate is well described by the following expression, 
\begin{equation}\label{E4}
\mathcal{R} \equiv \left<\mathcal{R}\right> ae^{-b(t/t_{12})^{c}},
\end{equation}
where $\left<\mathcal{R}\right>$ is the average merger rate over 12 Gyr, and $a$, $b$ and $c$ are the fitting parameters. $\mathcal{R}$ is defined as the number of mergers per unit time bin. 
Our best-fit parameters for the time evolution of the merger rate are ($a$, $b$, $c$) $=$ (13.01$\pm$3.00, 4.14$\pm$0.19, 0.35$\pm$0.04) for the IBP distribution. 
In order to test the dependence of the best-fit parameters on the cluster's initial number of stars, we repeated the fit for subsets of the simulation data with different initial number of stars ($N=2\times 10^5$, $5\times 10^5$ and $10^6$) and found that the time evolution of the normalized merger rates does not significantly depend on the initial number of stars in GCs. We also tested other subsets with different half-mass radii, galactocentric distances, binary fractions and we did not find any significant discrepancy from the best fitting parameters obtained for the entire survey of simulations.

\begin{figure}
  \centering
  \includegraphics[trim=15 10 5 5,width=1.0\columnwidth]{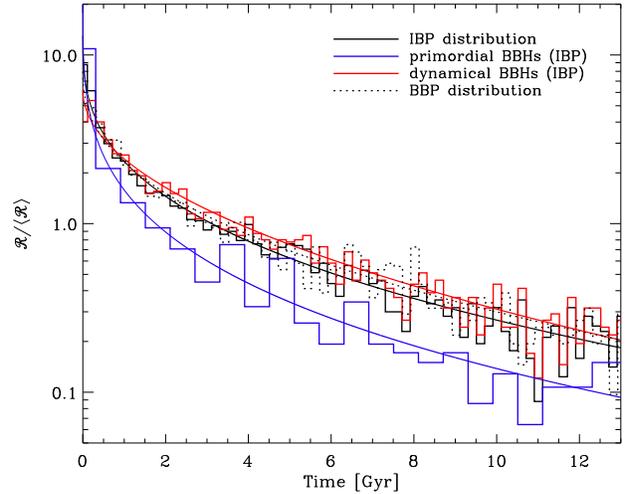}
  \caption{Same as Fig. \ref{F7} but the BBH mergers used for the fitting are separated into the primordial and dynamical BBH mergers. Dotted line indicates the time evolution of the merger rate for GC simulation models with the BBP distribution. }\label{F8}
\end{figure}

\begin{table}
  \begin{center}
  \caption{Best fitting results for empirical relations. ``pri'' and ``dyn'' denote the best-fit results using the primordial and dynamical BBH mergers separately. $\gamma_{\rm dyn,12}$ present the expected number of dynamical BBH mergers using current GC properties. Note that for the BBP distribution, the BBH mergers originating from GCs are mostly dynamical BBHs.}
  \begin{tabular}{c c c c c}
  \hline
  \hline
  binary & \multicolumn{4}{c}{$N_{\rm merg}$ vs. GC properties (Eq. \ref{E1}, \ref{E2}, \ref{E3})}\\  \cline{2-5}
  distribution &  & $A$ & $B$ & $\alpha$ \\
  \hline
  IBP & $\gamma_{\rm tot}$ & 12.53$\pm$0.22 & 6.89$\pm$0.84 & 0.33$\pm$0.02\\
  IBP & $\gamma_{\rm pri}$ &  - & 6.64$\pm$0.25 & -\\
  IBP & $\gamma_{\rm dyn}$ & 12.30$\pm$0.44 & - & 0.33$\pm$0.01\\
  IBP & $\gamma_{\rm dyn,12}$ & 303.1$\pm$6.1 & - & 0.43$\pm$0.02 \\
  BBP & $\gamma_{\rm tot}$ & 14.55$\pm$0.31 & -0.03$\pm$0.03 & 0.38$\pm$0.01\\
  BBP & $\gamma_{\rm dyn,12}$ & 527.7$\pm$11.1 &-& 0.52$\pm$0.02\\\hline\hline
  binary & \multicolumn{4}{c}{Time evolution of merger rate (Eq. \ref{E4})}\\  \cline{2-5}
  distribution &  & $a$ & $b$ & $c$ \\\hline
  IBP & $\mathcal{R_{\rm tot}}$ & 13.01$\pm$3.00 & 4.14$\pm$0.19 & 0.35$\pm$0.04\\
  IBP & $\mathcal{R_{\rm pri}}$ & 21.80$\pm$8.82 & 5.33$\pm$0.29 & 0.29$\pm$0.05 \\
  IBP & $\mathcal{R_{\rm dyn}}$ & 6.15$\pm$1.23 & 3.27$\pm$0.17 & 0.51$\pm$0.06\\
  BBP & $\mathcal{R_{\rm tot}}$ & 7.96$\pm$2.03 & 3.54$\pm$0.21 & 0.42$\pm$0.06\\\hline
  \label{T1}
  \end{tabular}
  \end{center}
\end{table}

In Fig. \ref{F8}, we present the time evolution of the merger rates for BBHs with different formation origins and find some differences between the  primordial and dynamical BBH merger rates. Since the progenitor BBHs for primordial mergers form in a very short time interval during a GC's early evolution ($T<30$ Myr), the merger rate decreases more rapidly than that for dynamical BBH mergers. The best-fit results are ($a$, $b$, $c$) $=$ (21.80$\pm$8.82, 5.33$\pm$0.29, 0.29$\pm$0.05) for primordial mergers and ($a$, $b$, $c$) $=$ (6.15$\pm$1.23, 3.27$\pm$0.17, 0.51$\pm$0.06) for dynamical mergers. For the BBP distribution, the best-fit parameters are ($a$, $b$, $c$) $=$ (7.96$\pm$2.03, 3.54$\pm$0.21, 0.42$\pm$0.06), similar to the parameters found for the dynamical BBH mergers in models with the IBP distribution.
We summarize our results for the correlation between $N_{\rm merg}$ and GC properties and the time evolution of the merger rates in Table \ref{T1}.

\subsection{Chirp mass and mass ratio of BBH mergers\label{S3.3}}
\begin{figure}
  \centering
  \includegraphics[trim=15 10 5 5,width=1.0\columnwidth]{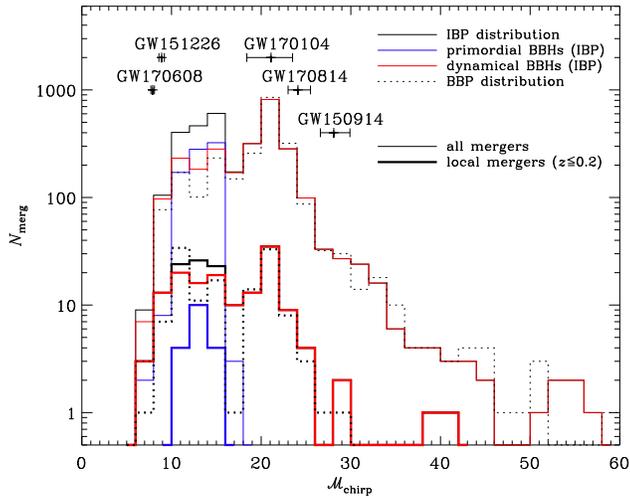}
  \caption{Distribution of chirp masses, $\mathcal{M}_{\rm chirp}$ of merging BBHs escaping from GC simulation models. Black line shows the distribution of all BBH mergers from GC simulation models with the IBP distribution. Blue and red lines are for primordial and dynamical BBH mergers, respectively. Dotted line shows $\mathcal{M}_{\rm chirp}$ distribution of BBH mergers from models with the BBP distribution. Thin and thick lines represent $\mathcal{M}_{\rm chirp}$ distribution for the BBH mergers throughout all look-back time and those merging in the local Universe ($z\leq0.2$), respectively. Note that the model lines correspond to merger rate densities and are not corrected for observational selection effects. The $\mathcal{M}_{\rm chirp}$ of 5 GW events that have been detected by LIGO so far are marked with the range of 90 per cent confidence intervals (data from \url{https://losc.ligo.org/events/}).  }\label{F9}
\end{figure}

In this section we investigate some fundamental properties of the merging BBHs. Fig. \ref{F9} shows the distribution of chirp masses, $\mathcal{M}_{\rm chirp}\equiv (m_1m_2)^{3/5}(m_1+m_2)^{-1/5}$ of merging BBHs. It is interesting to note that the chirp mass distribution for BBH mergers throughout all look-back time, $t_{\rm lb}$ and that for BBHs merging in the local Universe with red-shift, $z\leq0.2$ (i.e., $t_{\rm lb}\lesssim2.4$ Gyr with standard cosmological parameters assumed) are slightly different because the formation and merging timescales depend on the mass of BBHs \citep{2017ApJ...836L..26C,2018PhRvL.120o1101R}.
In Fig. \ref{F9} we also plot the values of the chirp masses of BBH mergers detected by LIGO so far and show that they approximately fall within the range of values corresponding to the broad peak in the  distribution of $\mathcal{M}_{\rm chirp}$ for merging BBHs. 
As illustrated by this figure, for the set of simulations considered in this paper the high-$\mathcal{M}_{\rm chirp}$ events are likely to belong to the dynamical BBH groups while the low-$\mathcal{M}_{\rm chirp}$ could be either primordial or dynamical BBHs. It is also possible that these low-$\mathcal{M}_{\rm chirp}$ sources come from metal-richer environments \citep[see e.g.][]{2017ApJ...836L..26C,2018arXiv180601285A}.

We point out that the distribution of $\mathcal{M}_{\rm chirp}$ of dynamical BBH mergers for the IBP distribution is almost identical to that for the BBP distribution (which has only dynamical BBH mergers). This implies that the host GC dynamics is the key factor determining the BBH mergers' properties and differences between the IBP and the BBP initial properties of primordial binaries do not play an important role in the chirp mass of the BBH mergers.
It is interesting to note that recent numerical studies \citep{2018PhRvL.120o1101R,2018ApJ...855..124S} for GCs with post-Newtonian calculations for BBHs suggested that mergers of BBHs can occur inside GCs and the merger product can form a binary with other BHs and merge again in/outside of clusters \citep[see also][for the retention of in-cluster BBH mergers]{2018arXiv180201192M}.
The mass distribution of merging BBHs, especially for high masses, will be affected by this process.
\begin{figure}
  \centering
  \includegraphics[trim=15 10 5 5,width=1.0\columnwidth]{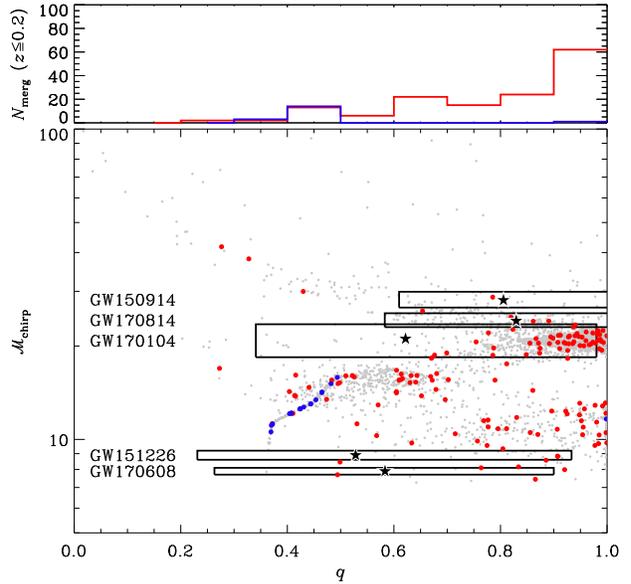}
  \caption{Distribution of BBH mergers for the IBP distribution in $q$-$\mathcal{M}_{\rm chirp}$ plane. Grey dots show all mergers and red and blue dots represent, respectively, dynamical and primordial BBHs that merge in the local Universe ($z\leq0.2$).
  Black stars and boxes indicate the ranges of $\mathcal{M}_{\rm chirp}$ and $q$ based on 90 per cent confidence intervals of LIGO detections. Upper panel shows the distribution of mass ratio, $q$, of primordial (blue) and dynamical (red) BBHs merging in the local Universe.}\label{F10}
\end{figure}

Fig. \ref{F10} shows the distribution of the $\mathcal{M}_{\rm chirp}$ versus the mass ratio $q$ ($\equiv m_{2}/m_{1}$, where $m_1>m_2$) of BBH mergers from models with the IBP distribution. 
The sequence of blue points corresponds to the primordial BBH mergers obtained with the binary stellar evolution and the mass fall-back mechanism used in the simulations (see Section \ref{S2}). Three overdense regions in this plane can be easily identified at  $\mathcal{M}_{\rm chirp}\sim20$ and $q\sim1$, $\mathcal{M}_{\rm chirp}\sim13$ and $q\sim0.55$, $\mathcal{M}_{\rm chirp}\sim10$ and $q\sim1$, respectively. This features are related to the shape of the mass function of BHs produced in our simulations, which has a bi-modal distribution with peaks at $m_{\rm BH}\sim12$ and 24 M$_{\odot}$.
In the upper panel in Fig. \ref{F10}, the distribution of $q$ shows that dynamical BBH mergers tend to have similar masses \citep[see also][]{2016MNRAS.458.3075A,2016PhRvD..93h4029R,2017MNRAS.469.4665P}.
It is interesting to point out the presence of a sequence of BBH mergers with high-$\mathcal{M}_{\rm chirp}$ and low-$q$. This group comprises BHs that may have increased their mass due to mergers with other stars or black holes. In the latter case, it may be possible that some of the merger remnants may already have been ejected from the host stellar system due to gravitational wave recoil kicks \citep{2018PhRvL.120o1101R,2018arXiv180201192M}.

An important general point to emphasize is that the $\mathcal{M}_{\rm chirp}$ distribution of merging BBHs strongly depends on the metallicity \citep{2016PhRvD..93h4029R,2017ApJ...836L..26C,2017MNRAS.464L..36A,2018MNRAS.474.2959G}. 
The metallicity affects not only the number of BHs produced, the number of BBH mergers but also the mass range of BBHs. 
However, \citet{2017ApJ...836L..26C} pointed out that the $\mathcal{M}_{\rm chirp}$ distribution of BBHs formed dynamically and merging in the local Universe ($z\leq0.2$) does not depend on metallicity for $Z\leq0.001$.

We note that the masses of BHs depend also on the single and binary stellar evolution recipes,
\footnote{Note that the common-envelope phase (CEP) is also important for the binary stellar evolution and the formation of compact binaries \citep[e.g.][]{2017MNRAS.468.2429B,2018arXiv180600001G}. In this study we used the CEP parameters, $\alpha_{\rm CE}=3$ and $\lambda=0.5$. However, recent studies \citep[e.g.][]{2017MNRAS.468.2429B} suggested $\alpha_{\rm CE}\sim 0.5$, and lower $\alpha_{\rm CE}$ and $\lambda$ value may lead to more binary mergers during the CEP and the subsequent production of single BHs. The uncertainty in the value of these parameters may affect the number of primordial BBHs and the mass distribution of merging BBHs from GCs.}
and, in particular, on the fall-back prescription \citep[e.g.][]{2012ApJ...749...91F,2015MNRAS.451.4086S}. Additional observations and numerical simulations are therefore needed to constrain the values of the BH masses after supernova explosions.

\section{Merger rate density\label{S4}}
In this section we estimate the merger rate density of BBHs escaping from the GCs using the empirical relations obtained in the previous sections. Since the relation between the expected number, the mass distribution and the time evolution of the rates are different for primordial and dynamical BBH mergers, we calculate the merger rates separately for BBH mergers with different origins.

\subsection{Rate density for primordial BBH mergers\label{S4.1}}
To estimate the merger rate density, we follow the calculation of  \citet{2017MNRAS.464L..36A} \citep[see also][]{2004A&A...415..407B}. 
For this calculation, we use the merging time and the chirp mass for each primordial BBH that will merge within 12 Gyr from all the simulation models. Having this data and the total and average initial mass of all simulated GCs, we can estimate the merger rate density per unit chirp mass using a GC star formation rate as a function of redshift and the contribution of the merger rate from individual GCs to the rate density according to the age distribution of GCs based on the GC star formation history. For this purpose, the GC star formation rate estimated by \citet{2013MNRAS.432.3250K} has been adopted in this calculation. 

We already pointed out that the number of primordial BBH mergers depends on the initial mass and binary fraction of GCs. If we simply assume that the initial binary fraction is universal for all GCs, the merger rate density for primordial BBHs only depends on the GC formation rate.
The number of primordial BBH mergers over 12 Gyr based on the IBP distribution is $\sim$6.64$f_{\rm b,0}$ per $10^5$M$_{\odot}$ from Eq. (\ref{E3}) and its best-fit parameters. On the other hand the number of primordial BBH mergers from the simulations with the BBP distribution is very small and we estimate the contribution of primordial BBH mergers to be negligible in this case.
Our estimate of the number of mergers per unit mass is consistent with that in \citet{2018MNRAS.474.2959G} with similar metallicity. From this number it follows that the local merger rate density of primordial BBHs ranges from 0.18 to 1.8 ${\rm Gpc}^{-3}{\rm yr}^{-1}$  for an initial binary fraction ranging from 10 to 100 per cent.

\subsection{Rate density for dynamical BBH mergers from initial GC properties\label{S4.2}}
In order to estimate the merger rate density for dynamical BBH mergers from our empirical relations, we first need to calculate the number of BBH mergers per unit GC mass. For the calculation of the number of dynamical BBH mergers we need to make an assumption on the initial GC mass and size distribution and then combine these with our estimate of the number of dynamical BBH mergers from Eq. (\ref{E2}). For the initial GC mass function (ICMF), we adopt a \citet{1976ApJ...203..297S} function
\begin{equation}
    dN_{\rm GC}\propto M^{-\beta}{\rm exp}(-M/M_{*})dM
\end{equation}
where $\beta=2$ \citep{1999ApJ...527L..81Z,2003A&A...397..473B,2003AJ....126.1836H} and $M_{*}$ is the exponential cut-off mass of the ICMF.
We consider different combinations of the minimum mass of GCs, $M_{\rm min}=10^3, 10^4$M$_{\odot}$ and exponential cut-off mass $M_{*}=10^6, 10^{6.5}$M$_{\odot}$ \citep[for the selection of $M_{*}$, see e.g.][]{2017ApJ...839...78J}. 

No firm prediction on the distribution of the initial sizes of GCs is currently available. Instead, we tried to find a realistic distribution of the initial size of GCs from the observations of young massive clusters (YMCs) in extra-galactic systems although it is possible that old GCs forming in the early Universe formed with a different size distribution. There are a number of observational studies \citep[e.g.][]{2010ApJ...709..411H,2012MNRAS.419.2606B,2015MNRAS.452..525R} showing that the effective radii of YMCs tend to increase with the YMC's age. This might be due to the combined effects of the primordial gas expulsion, initial mass loss by the stellar evolution and/or the presence of a significant number of retained BHs \citep[e.g.][]{2008MNRAS.386...65M}. By correcting the age dependence of the effective radii of YMCs in M83 \citep{2015MNRAS.452..525R}, we obtain a log-normal distribution of the initial half-mass radius with $\sigma=0.4$ and $\left<r_{\rm h}\right>=2.8$ pc which is comparable with the initial half-mass radii used in the numerical simulations by \citet{2010ApJ...719..915C,2013MNRAS.429.2881C} reproducing the distribution of GGCs.

Many studies of YMCs found that there is a weak correlation between the mass and the effective radius of YMCs \citep[e.g.][]{1999AJ....118..752Z,2004A&A...416..537L,2010ApJ...709..411H,2012A&A...543A...8M,2017ApJ...841...92R}. We take the relation for the average value of the initial half-mass radius, $\left<r_{\rm h}\right>/{\rm pc}=2.8\times(M/10^4{\rm M}_{\odot})^{0.1}$ from \citet{2004A&A...416..537L}. In order to investigate the effects of the initial size distribution on the merger rate density, we consider another distribution of the initial half-mass radius, $\left<r_{\rm h}\right>/{\rm pc}=0.33\times(M/10^4{\rm M}_{\odot})^{0.13}$ from \citet{2012A&A...543A...8M}, which is much smaller that the previous one \citep[note that these ``small'' and ``large'' size distributions are roughly consistent with the half-mass radii for massive clusters and open clusters/associations from the simulations for the cluster formation done by][]{2016ApJ...817....4F}. We, however, ignore the effects of the host galaxy tidal field on the initial distribution of half-mass radii since there is no correlation between the effective radii and galactocentric distances of YMCs found in nearby galaxies \citep{2017ApJ...841...92R}. \citet{2012ApJ...756..167M} also showed that the galactocentric distance does not significantly affect the early (less than a few hundreds Myr) evolution of half-mass radii of star clusters. 

Using the initial mass and size distributions discussed above, 
we can estimate the expected number of dynamical BBH mergers per GC masses through Eq. (\ref{E2}) and its fitting parameters as 
\begin{equation}
    \frac{N_{\rm merg}}{M_{\rm GCSF}}=\frac{\iint \gamma_{\rm dyn} N(M)N(r_{\rm h})dMdr_{\rm h}}{\int N(M)MdM},
\end{equation}
where $M_{\rm GCSF}$ is the total mass of all GCs, and $N(M)$ and $N(r_{\rm h})$ are, respectively, the mass and half-mass radius distribution of initial GC systems (and in which, as explained above, the mean of the half-mass radius distribution depends on the cluster mass).
For the different ICMF we have considered, we find that the total number of dynamical BBHs mergers per unit mass over 12 Gyr based on the ``large'' size distribution \citep{2004A&A...416..537L} is
$\sim$2.45 (2.34) per $10^5$M$_{\odot}$ for [$M_{\rm min}, M_{*}$] = [$10^4$M$_{\odot}, 10^{6.5}$M$_{\odot}$], $\sim$2.11 (1.96) for [$10^4$M$_{\odot}, 10^{6}$M$_{\odot}$], $\sim$1.96 (1.82) for [$10^3$M$_{\odot}, 10^{6.5}$M$_{\odot}$], and $\sim$1.69 (1.53) for [$10^3$M$_{\odot}, 10^{6}$M$_{\odot}$], respectively for the simulation models with the IBP (BBP) distribution. 

From these estimates, the corresponding local merger rate densities for the different ICMF are 1.91 (1.73), 1.64 (1.45), 1.52 (1.34) and 1.31 (1.13) ${\rm Gpc}^{-3}{\rm yr}^{-1}$, respectively. 
We point out that the ratio of the local merger rate density to the number of mergers per GC masses for dynamical BBHs is larger than that for primordial BBHs due to the chirp mass distribution and time evolution of the merger rates (see Figs. \ref{F8} and \ref{F9}). 

We also emphasize that the local merger rate density for dynamically-formed BBHs does not show any significant dependence on the binary distributions and weakly dependent on the ICMF with a variation for the different ICMFs considered of a factor of $\lesssim2$. Along with the local merger rate density for primordial BBHs, our calculation of local merger rate density of $\sim$4 ${\rm Gpc}^{-3}{\rm yr}^{-1}$ is consistent with that of $\sim$5 ${\rm Gpc}^{-3}{\rm yr}^{-1}$ from other literature \citep[e.g.][]{2016PhRvD..93h4029R,2017MNRAS.464L..36A}. Some discrepancies may be due to the different distribution of GC models.

For more compact initial size distribution of GCs suggested by \citet{2012A&A...543A...8M}, we obtain the merger rate density of 14.3 (17.5), 12.6 (15.1), 11.7 (13.9) and 10.3 (12.0) ${\rm Gpc}^{-3}{\rm yr}^{-1}$, respectively for GC models with the IBP (BBP) distribution with the different ICMF. The merger rate density is larger for the BBP distribution because the expected number of dynamical BBH mergers has stronger correlation with the initial density of GCs. The many more additional detections of BBH mergers will be needed to shed light on the initial binary distribution in GCs as well as the distribution of the initial properties of GCs.

We point out that 
the mass distribution and the merger rate are nearly independent of the metallicity for $Z\leq0.001$ for either primordial \citep{2018MNRAS.474.2959G} or dynamical \citep{2017ApJ...836L..26C} BBH mergers. By combining the GC star formation history from \citet{2013MNRAS.432.3250K} and the red-shift metallicity relation from \citet{2016Natur.534..512B}, we expect that approximately $\lesssim$10\% of GCs especially forming at lower red-shift ($z\sim2$--3) will be affected by the effects of metallicity. Although we fixed the metallicity to $Z=0.001$, there will not be significant effects of cosmological metallicity variation on the estimation of the local merger rate density.

\subsection{Rate density for dynamical BBH mergers from current GC properties\label{S4.3}}
In Section \ref{S3.1}, we discussed the correlation between the expected number of dynamical BBH mergers and the current GCs' mass and half-mass density. Combining the correlation and the distribution of observed GCs, we also can estimate the merger rate density for BBH mergers originating from surviving GCs. To reproduce the mass distribution of GGCs, we generated the GCMF following an evolved Schechter function \citep{2007ApJS..171..101J},
\begin{equation}
\frac{dN}{dM}\propto \frac{1}{(M+\Delta)^2}{\rm exp} \left( -\frac{M+\Delta}{M_{\rm c}}\right),
\end{equation}
where $\Delta$ is a factor for the mass loss of GCs, and $M_{\rm c}$ is the exponential cut-off mass for the GCMF. We adopted the values of $\Delta=10^{5.4}$M$_{\odot}$ and $M_{\rm c}=10^{5.9}$M$_{\odot}$ from \citet{2007ApJS..171..101J}.

\begin{figure}
  \centering
  \includegraphics[trim=15 10 5 5,width=1.0\columnwidth]{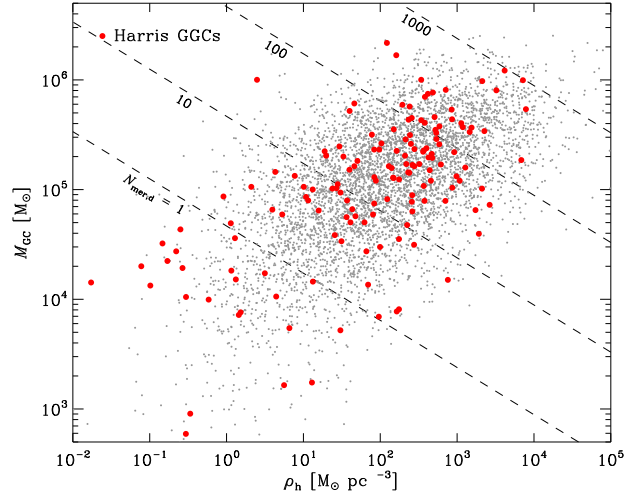}
  \caption{Present-day distribution of GCs in the half-mass density and the mass of GCs. Red dots are the Milky Way GCs from the \citet{1996AJ....112.1487H} GGC catalog assuming that the mass-to-light ratio for all GCs is equal to 2 and the half-mass radius is equal to $\sim$1.7 of the projected half-light radius in the catalog. Grey dots are modeled GCs following an evolved Schechter function GCMF \citep{2007ApJS..171..101J} and a log-normal distribution with $\left<r_{\rm h}\right>=6.1$ pc and $\sigma=0.63$ for the distribution of the half-mass radius. Dashed lines indicate the expected number of dynamical BBH mergers produced in the individual GCs over 12 Gyr lifetime. }\label{F11}
\end{figure}

In Fig. \ref{F11}, we show the distribution of GGCs from \citet{1996AJ....112.1487H} catalog in the $M_{\rm GC}$-$\rho_{\rm h}$ plane. We simply assumed that the mass-to-light ratio $\Upsilon=2$ and the half-mass radius $r_{\rm h}= \sim 1.7 R_{\rm hl}$ \citep[where $R_{\rm hl}$ is the projected half-light radius, see ][]{2013MNRAS.429.2881C} for all GGCs.
We then distributed the half-mass radius of GCs at present-day by using a log-normal distribution with parameters $\left<r_{\rm h}\right>=6.1$ pc and $\sigma=0.63$ which give a best-fit with the distribution of \citet{1996AJ....112.1487H} GGCs in the $M_{\rm GC}$-$\rho_{\rm h}$ plane. 
Dashed lines in this figure indicate the expected number of dynamical BBH mergers generated from the individual GCs over 12 Gyr cluster lifetime. Many of GGCs are expected to produce between $\sim$10 and $\sim$ 1000 BBH mergers within 12 Gyr. 

\begin{table}
  \begin{center}
  \caption{Estimate of the local merger rate density, $R_{\rm local}$. We used the GC star formation rate from \citet{2013MNRAS.432.3250K} for the calculation of $R_{\rm local}$ of primordial and dynamical BBHs based on the initial GC properties. We used a \citet{1976ApJ...203..297S} function with different parameters for the ICMF. ``large'' and ``small'' denote the initial distribution of the half-mass radii of GCs based on \citet{2004A&A...416..537L} and \citet{2012A&A...543A...8M}, respectively. For the GCMF for the calculation of $R_{\rm local}$ from current GC properties, we used an evolved Schechter \citep{2007ApJS..171..101J} function and a log-normal distribution used in \citet{2016PhRvD..93h4029R}. A log-normal distribution for the initial size ($r_{\rm h}$) distribution of GCs was used. For $\rho_{\rm GC}$, we took conservative, standard and optimistic cases from \citet{2016PhRvD..93h4029R}. We also considered a time-dependent $\rho_{\rm GC}$ from \citet{2017PASJ...69...94F}, as denoted by ``F2017''. See the text for more details.}
  \begin{tabular}{c c c c c c}
  \hline
  \hline
  \multicolumn{6}{c}{Primordial BBH mergers (Section \ref{S4.1})}\\
  \hline
  $f_{\rm b,0}$ & \multicolumn{5}{c}{$R_{\rm local}$ (Gpc$^{-3}$yr$^{-1}$)}\\
  \hline
  10\% & \multicolumn{5}{c}{0.18} \\
  100\% & \multicolumn{5}{c}{1.8}\\
  \hline
  \hline
  \multicolumn{6}{c}{Dynamical BBH mergers w/ initial GC properties (Section \ref{S4.2})}\\
  \hline
  binary & ICMF & \multicolumn{4}{c}{initial GC size distribution} \\\cline{3-6}
  distribution & [$M_{\rm min}, M_{*}$] & \multicolumn{2}{c}{large} & \multicolumn{2}{c}{small}\\\hline
  IBP & [$10^4, 10^{6.5}$] M$_{\odot}$ & \multicolumn{2}{c}{1.91} & \multicolumn{2}{c}{14.3}\\
  IBP & [$10^4, 10^{6}$] M$_{\odot}$ & \multicolumn{2}{c}{1.64} & \multicolumn{2}{c}{12.6}\\
  IBP & [$10^3, 10^{6.5}$] M$_{\odot}$ & \multicolumn{2}{c}{1.52} & \multicolumn{2}{c}{11.7}\\
  IBP & [$10^3, 10^{6}$] M$_{\odot}$ & \multicolumn{2}{c}{1.31} & \multicolumn{2}{c}{10.3}\\
  BBP & [$10^4, 10^{6.5}$] M$_{\odot}$ & \multicolumn{2}{c}{1.73} & \multicolumn{2}{c}{17.5}\\
  BBP & [$10^4, 10^{6}$] M$_{\odot}$ & \multicolumn{2}{c}{1.45} & \multicolumn{2}{c}{15.1}\\
  BBP & [$10^3, 10^{6.5}$] M$_{\odot}$ & \multicolumn{2}{c}{1.34} & \multicolumn{2}{c}{13.9}\\
  BBP & [$10^3, 10^{6}$] M$_{\odot}$ & \multicolumn{2}{c}{1.31} & \multicolumn{2}{c}{12.0}\\
  \hline
  \hline
  \multicolumn{6}{c}{Dynamical BBH mergers w/ current GC properties (Section \ref{S4.3})}\\
  \hline
  binary & GCMF & \multicolumn{4}{c}{$\rho_{\rm GC}$ (Mpc$^{-3}$)}\\\cline{3-6}
  distribution &  & 0.33 & 0.77 & 2.31 & F2017\\\hline
  IBP & eSchechter & 0.63 & 1.46 & 4.39 & 5.16 \\
  IBP & log-normal & 0.89 & 2.02 & 6.20 & 7.28 \\
  BBP & eSchechter & 0.84 & 1.96 & 5.88 & 6.77 \\
  BBP & log-normal & 1.15 & 2.68 & 8.04 & 9.26 \\\hline
  \label{T2}
  \end{tabular}
  \end{center}
\end{table}

To compute the local merger rate density from the expected number of BBH mergers from current GCs properties, $\gamma_{\rm dyn,12}$ from Eq. (\ref{E2}) and the distributions of GCs' present-day properties introduced above, we need the number density of GCs, $\rho_{\rm GC}$ in the local Universe. We simply adopt 0.33, 0.77 and 2.31 Mpc$^{-3}$ \citep{2016PhRvD..93h4029R} for conservative, standard and optimistic assumptions of $\rho_{\rm GC}$, respectively. We obtain the local merger rate densities for dynamical BBH mergers are 0.63, 1.46 and 4.39 ${\rm Gpc}^{-3}{\rm yr}^{-1}$ for conservative, standard and optimistic cases, respectively, assuming the age of all GCs is 12 Gyr. The merger rate at the present-day has been corrected by using the Eq. (\ref{E4}) for the case of dynamical BBH mergers (i.e. $\mathcal{R} \sim 0.24 \left<\mathcal{R}\right>$).

We have also considered the time-dependent $\rho_{\rm GC}$ from \citet{2017PASJ...69...94F} for old (T $\ge$ 10 Gyr) GCs. The total $\rho_{\rm GC}$ is slightly smaller (2.2 Mpc$^{-3}$) than the optimistic case of \citet{2016PhRvD..93h4029R}. However, we obtain the merger rate density of 5.16 ${\rm Gpc}^{-3}{\rm yr}^{-1}$ which is slightly larger than our optimistic case based on $\rho_{\rm GC}$ from \citet{2016PhRvD..93h4029R} because there are younger GCs with higher merger rates compared to GCs with ages of 12 Gyr (see Figs. \ref{F7} and \ref{F8}). Our estimates are similar in order of magnitude but systematically smaller than those from other studies such as $\sim$5 ${\rm Gpc}^{-3}{\rm yr}^{-1}$ from \citet{2016PhRvD..93h4029R} (for the standard case) and 13 ${\rm Gpc}^{-3}{\rm yr}^{-1}$ from \citet{2017PASJ...69...94F} \citep[see also][]{2017MNRAS.464L..36A,2017MNRAS.469.4665P} 
since our GCMF includes a larger number of GCs with lower masses which contribute less to the merger rate density compared to the more massive GCs.
Using the same GCMF adopted in \citet{2016PhRvD..93h4029R}, we obtain the merger rate density of 0.89, 2.02, 6.20 and 7.28 ${\rm Gpc}^{-3}{\rm yr}^{-1}$ for conservative, standard, optimistic and time-dependent $\rho_{\rm GC}$. 

On the other hand, as discussed in previous sections, most of BBH mergers based on the BBP distribution are dynamical mergers. We estimate the merger rate density of 0.84, 1.96, 5.88 and 6.77 ${\rm Gpc}^{-3}{\rm yr}^{-1}$ for conservative, standard, optimistic and time-dependent $\rho_{\rm GC}$, respectively, using the evolved Schechter function GCMF. The estimate of the merger rate density becomes 1.15, 2.68, 8.04 and 9.26 assuming that the GCMF follows a log-normal distribution as used in \citet{2016PhRvD..93h4029R}. We summarize our estimates of the local merger rate density in Table \ref{T2}.

We point out that the calculation of the local merger rate density based on the cluster current properties includes only the contribution of surviving clusters.
GCs dissolving before 12 Gyr of course can contribute to the population of BBH mergers and to take their contribution into account, a calculation like that presented in the previous section must be carried out.

\begin{figure}
  \centering
  \includegraphics[trim=15 10 5 5,width=1.0\columnwidth]{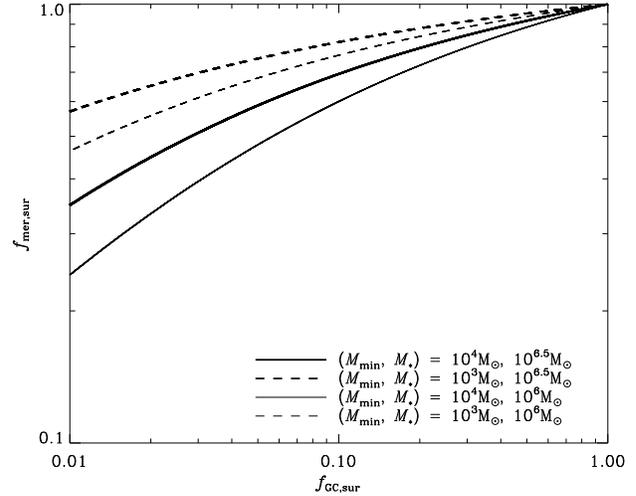}
  \caption{Cumulative fraction of dynamical BBH mergers from surviving GCs as a function of the fraction of surviving GCs. We simply assume that GCs are disrupted from lower-mass GCs. Different lines present the ICMF with different parameters.}\label{F12}
\end{figure}

Alternatively we can use a simple toy model for the evolution of a globular cluster system and assume the GC disruption proceeding from the low-mass GCs; using this simple model we can calculate the cumulative fraction of BBH mergers from surviving GCs as a function of the fraction of surviving GCs from our GC models introduced in Section \ref{S4.2}.
We show the result of this calculation in Fig. \ref{F12}. This figure provides an approximate estimate of the fraction of BBH mergers from GCs that still survive at the present-day. If we assume that only $\sim$3 per cent of GCs survive up to now as suggested by \citet{2014ApJ...785...71G} for the Milky Way, the fraction of BBH mergers from surviving GCs varies from 0.3 to 0.7 \citep[0.5 with $M_{\rm min}=10^4 {\rm M}_{\odot}$ used in][]{2014ApJ...785...71G} depending on the ICMF.
The local merger rates based on the current GC properties become comparable with those based on the initial GC properties with ``small'' size distribution when the contribution of merging BBHs from dissolving GCs is taken into account.

We conclude this section by pointing out that it is possible that YMCs can contribute to the local merger rate density. 
\citet{2017MNRAS.467..524B,2018MNRAS.473..909B} have performed direct $N$-body simulations for YMC-like systems with post-Newtonian approximation implemented and found that YMCs can contribute the detection rate to a similar extent as more massive GC counterpart. \citet{2014MNRAS.441.3703Z} have estimated the local merger rate density of 3.6 Gpc$^{-3}$yr$^{-1}$  for BBHs originating from YMCs in the local Universe. 
\citet{2017PASJ...69...94F} also have suggested from their direct $N$-body simulations that the local merger rate density can be up to a factor of $\sim$3 times larger when younger clusters with ages between 2 and 10 Gyr are included in the estimation of the merger rate density. Using the empirical relation for the time evolution of the merger rates in Eq. (\ref{E4}) and the GC formation rate adopted by \citet{2017PASJ...69...94F}, we obtain a local merger rate density about $\sim$6 times higher when younger GCs are included.
When younger clusters are included, an important aspect to consider is the well-known age-metallicity relation for GCs \citep[e.g.][]{2013MNRAS.436..122L}. \citet{2017PASJ...69...94F} considered the effects of the metallicity by limiting the mass of BHs and found that there is no significant effect on the local merger rates. This is, however, in contrast with the findings of \citet{2018MNRAS.474.2959G} who suggested that the number of BBH mergers per unit mass strongly depends on the metallicity. The study of \citet{2018MNRAS.474.2959G} is focused on primordial BBHs, but in the dense environment like GCs, the internal dynamics can in part compensate the effects of the metallicity \citep[see e.g.][]{2017MNRAS.464L..36A,2017ApJ...836L..26C}.

Finally in this study, we did not consider the contribution to the merger rate density by the BBHs that merge inside GCs through the dynamical interactions and binary evolution. However, these in-cluster mergers become more important for very young clusters \citep{2017MNRAS.464L..36A,2017MNRAS.467..524B,2018MNRAS.473..909B}. According to \citet{2017MNRAS.464L..36A}, the contribution of these in-cluster mergers is about 20 per cent of the total merger rate through the entire evolution however becomes less than 1 per cent if the host GCs are old (T $>$ 10 Gyr). It is also important to note that the rate of in-cluster mergers can increase substantially if the dissipative effects connected with GW radiation (i.e. three-body GW capture) are taken into account \citep{2018ApJ...855..124S}.

\section{summary and conclusions\label{S5}}
In this paper we have studied the formation of binary black holes (BBHs) in globular clusters (GCs) and explored the relation between the number and properties of merging BBHs and the structural properties of their host GCs. Our study is based on a large survey of Monte Carlo simulations following the dynamical evolution of GCs with a broad range of different initial masses, sizes and primordial binary properties.

Our results have revealed a close correlation between the number of BBH mergers escaping from GCs and the properties of host GCs such as the initial mass, half-mass radius and the fraction of primordial binaries (Figs. \ref{F1} and \ref{F4}).

We identified two groups of BBH mergers; one group is composed of primordial BBH mergers forming simply as a result of binary stellar evolution and escaping from GCs due to the natal kicks by supernova explosions. The second group is composed of  dynamical BBH mergers forming as a result of binary-binary and binary-single interactions in the GC dense environments and ejected from GCs through the dynamical interactions.

The number of primordial BBH mergers is correlated with the GC's initial mass and binary fraction (see Eq. \ref{E3}), while we found that the number of dynamical BBH mergers produced in 12 Gyr is correlated with a parameter $\gamma_{\rm dyn}$ (see Eq. \ref{E2}) depending on the GC's initial mass and half-mass density (Figs. \ref{F2} and \ref{F3}). Interestingly we have shown that the number of dynamical BBH mergers correlates also with the same $\gamma_{\rm dyn}$ parameter but defined in terms of the GC's current properties (Figs. \ref{F5} and \ref{F6}). We provide analytic expressions describing the correlations between the number of BBH mergers and the host GC's properties and apply them to estimate the BBH merger rate for a few different models of GC populations but the expression provided in our study can be used more in general for GC populations with initial conditions different from those adopted in our calculations.

The specific properties of primordial and dynamical BBH mergers such as the merging time and the chirp mass distribution are very important for the estimate of the local merger rate and the detection rate. In general, we find that the merger rate decreases with time due to the continuous ejection of single and binary BHs from GCs (Fig. \ref{F7}). We showed that the time evolution of the merger rate for primordial BBH mergers decreases more rapidly than that for dynamical BBH mergers; this difference is due to differences between the formation and ejection timescales of the two groups of BBH mergers (Fig. \ref{F8}).
The two groups of BBH mergers are characterized also by differences in the chirp masses.
The dynamical BBH mergers contribute more massive BBH mergers compared to the primordial BBH mergers (Figs. \ref{F9} and \ref{F10}).

Based on the analytic expressions obtained from study, we estimated the local merger rates of BBHs escaping from GCs. The local merger rate for primordial BBHs depends only on the cosmological GC formation rate and we obtained a rate of 0.18--1.8 ${\rm Gpc}^{-3}{\rm yr}^{-1}$ (Section \ref{S4.1}) depending on the primordial binary fraction.
To estimate the local merger rate for dynamical BBHs, on the other hand, it is necessary make an assumption on the initial distribution of GC masses and size. As pointed out above, the analytic expressions obtained in this paper allow to calculate the local merger rate for any assumption concerning these initial distributions. We estimated a local merger rate for dynamical BBHs of 1.3--18 ${\rm Gpc}^{-3}{\rm yr}^{-1}$ depending on a variety of combinations of the initial GC mass function and size distribution (Section \ref{S4.2}).
We also estimated a local rate for dynamical BBH mergers from the current properties of surviving GCs equal to 0.6--9.3 ${\rm Gpc}^{-3}{\rm yr}^{-1}$ (Section \ref{S4.3}; see also Table \ref{T2}), assuming all GCs have the same age and metallicity.

The production of BBH mergers from GCs also can be influenced by the formation and the presence of intermediate mass black holes (IMBHs) in GCs. \citet{2015MNRAS.454.3150G} suggested that a seed BH for an IMBH can be formed by the runaway collisions of massive main-sequence (MS) stars \citep[see also][]{2002ApJ...576..899P,2017MNRAS.472.1677S}. This process will preferentially deplete the massive MS progenitors for stellar-mass BHs. Moreover, \citet{2007MNRAS.374..857T} have found that hard binaries can be disrupted by the interactions with the IMBH. These interactions between the IMBH and BBHs might result in the capture of one BH to the IMBH and the ejection of the companion BH \citep[this IMBH-BH binary can deplete the stellar-mass BH population by ejection; see][]{2014MNRAS.444...29L}, which is the possible source of intermediate mass ratio inspirals (IMRIs) for space-based GW detectors \citep[e.g.][]{2002ApJ...581..438M,2009ApJ...698L.129S}. Detailed investigations for the effects of the formation of IMBHs in GCs on the merger rate of stellar-mass BBHs will be studied in our forthcoming papers.

\section*{acknowledgement}
We thank an anonymous referee whose suggestions helped to improve this manuscript.
JH acknowledges support from the China Postdoctoral Science Foundation, Grant No. 2017M610694.
AA was partially supported by the National Science Center (NCN), Poland, through the grant UMO-2015/17/N/ST9/02573.
MG and AA were partially supported by NCN, Poland, through the grant UMO-2016/23/B/ST9/02732 and is currently supported by the Carl Tryggers Foundation through the grant CTS 17:113.
TB was supported by the grant TEAM/2016-3/19 from the Foundation for Polish Science (FNP).
This research was supported in part by Lilly Endowment, Inc., through its support for the Indiana University Pervasive Technology Institute, and in part by the Indiana METACyt Initiative. The Indiana METACyt Initiative at IU is also supported in part by Lilly Endowment, Inc. This work benefited from support by the International
Space Science Institute (ISSI), Bern, Switzerland, through its International
Team programme ref. no. 393 \textit{The Evolution of Rich Stellar Populations
\& BH Binaries} (2017-18).

\end{document}